\documentclass[lettersize,journal]{IEEEtran}
\usepackage{amsmath,amsfonts}
\usepackage{algorithmic}
\usepackage{algorithm}
\usepackage{array}
\usepackage[caption=false,font=normalsize,labelfont=scriptsize,textfont=scriptsize]{subfig}
\usepackage{textcomp}
\usepackage{stfloats}
\usepackage{url}
\usepackage{verbatim}
\usepackage{graphicx}
\usepackage{cite}
\usepackage{multirow}
\usepackage{setspace}
\begin{document}

\title{\huge Toward Beamfocusing-Aided Near-Field Communications: Research Advances, Potential, and Challenges}
\author{Jiancheng An, \emph{Member, IEEE}, Chau Yuen, \emph{Fellow, IEEE}, Linglong Dai, \emph{Fellow, IEEE},\\Marco Di Renzo, \emph{Fellow, IEEE}, M\'erouane Debbah, \emph{Fellow, IEEE}, and Lajos Hanzo, \emph{Life Fellow, IEEE}
\thanks{J. An and C. Yuen are with the School of Electrical and Electronics Engineering, Nanyang Technological University, Singapore 639798 (e-mail: jiancheng\_an@sutd.edu.sg; chau.yuen@ntu.edu.sg). L. Dai is with the Beijing National Research Center for Information Science and Technology and the Department of Electronic Engineering, Tsinghua University, Beijing 100084, China (e-mail: daill@tsinghua.edu.cn). M. Di Renzo is with Universit\'e Paris-Saclay, CNRS, CentraleSup\'elec, Laboratoire des Signaux et Syst\`emes, 91192 Gif-sur-Yvette, France (e-mail: marco.di-renzo@universite-paris-saclay.fr). M. Debbah is with Khalifa University of Science and Technology, P O Box 127788, Abu Dhabi, UAE (e-mail: merouane.debbah@ku.ac.ae). L. Hanzo is with the School of Electronics and Computer Science, University of Southampton, SO17 1BJ Southampton, U.K. (e-mail: lh@ecs.soton.ac.uk).}\vspace{-0.6cm}}

\markboth{DRAFT}{DRAFT}

\maketitle

\begin{abstract}
Next-generation mobile networks promise to support high throughput, massive connectivity, and improved energy efficiency. To achieve these ambitious goals, extremely large-scale antenna arrays (ELAAs) and terahertz communications constitute a pair of promising technologies. This will result in future wireless communications occurring in the near-field regions. To accurately portray the channel characteristics of near-field wireless propagation, spherical wavefront-based models are required and present both opportunities as well as challenges. Following the basics of near-field communications (NFC), we contrast it to conventional far-field communications. Moreover, we cover the key challenges of NFC, including its channel modeling and estimation, near-field beamfocusing, as well as hardware design. Our numerical results demonstrate the potential of NFC in improving the spatial multiplexing gain and positioning accuracy. Finally, a suite of open issues are identified for motivating future research.
\end{abstract}

\section{Introduction}
With the widespread deployment of fifth-generation (5G) base stations (BSs), the exploration of new technologies for sixth-generation (6G) mobile networks has begun. Recently, the International Telecommunication Union Radiocommunication Sector (ITU-R) has defined six major usage scenarios for 6G with stringent requirements for fifteen diverse capabilities including data rate, connection density, latency, reliability, coverage, positioning, etc. To meet these ambitious goals, the use of extremely large-scale antenna arrays (ELAAs) at the BSs is envisioned as one of the promising solutions \cite{OJCS_2023_Elbir_Elbir_7, CM_2023_Cui_Near_56}. Additionally, abundant bandwidth resources can be found at higher carrier frequencies in the terahertz (THz) band \cite{arXiv_2023_Liu_Near_2}. Serendipity has it that these two technologies can benefit each other. Explicitly, the sub-millimeter wavelength allows packing an immense number of antenna elements (AEs) into a compact implementation, while the ELAAs provide substantial array gain to compensate for the severe path loss of the THz band.

As the array apertures become large and operate at extremely high frequencies, the radiating near-field range will considerably expand. Specifically, the Rayleigh distance, a commonly used metric to distinguish the near- and far-field regions, is proportional to both the square of the array aperture and to the radio frequency \cite{CM_2023_Cui_Near_56, arXiv_2023_Liu_Near_2}. For instance, for a $1 \text{m} \times 1 \text{m}$ square array operating at $28$ GHz, the Rayleigh distance is up to $373$ meters (m), essentially covering a microcell. Hence, near-field communications (NFC) will become the norm as ELAAs are incorporated into 6G mobile networks. To accurately model the channel characteristics in NFC, the general spherical wave-propagation model must be considered, invalidating the uniform planar wave approximation utilized in conventional far-field communication (FFC) systems \cite{TWC_2022_Lu_Communicating}.

Near-field propagation opens up appealing opportunities but also presents unique challenges for communication system design. \emph{Firstly}, the existing channel estimation and transmission schemes based on the far-field assumption are no longer applicable in NFC systems \cite{TWC_2022_Zhang_Beam_73, TCOM_2022_Cui_Channel_107, TWC_2023_Cui_Near}. \emph{Secondly}, implementing ELAAs is technically challenging due to the large number of AEs and bandwidth involved \cite{arXiv_2023_Liu_Near_2}. On the other hand, leveraging ELAAs has demonstrated the potential to focus the intended signals toward a specific spatial point \cite{TWC_2022_Zhang_Beam_73}. This focal selectivity is capable of mitigating the interference among users even when they have similar angles of incidence with respect to the ELAA. Additionally, NFC is expected to improve the spatial multiplexing gain of MIMO systems under line-of-sight (LoS) propagation conditions \cite{JSAC_2020_Dardari_Communicating_133}.

In this article, we provide an overview of NFC\footnote{This article focuses only on systems with extremely \emph{large} arrays, which differs from extremely \emph{dense} arrays known as holographic MIMO \cite{CL_2023_An_A}.}. Specifically, we commence by outlining the essential distinctions between NFC and FFC systems. We then discuss the key challenges in NFC system design and review the recent technological advances in both channel modeling and estimation, beamfocusing schemes, as well as hardware design. Furthermore, numerical results demonstrate two promising applications of NFC: \emph{i)} Increasing the MIMO multiplexing gain under LoS conditions and \emph{ii)} Enhancing the positioning accuracy of wireless networks. Finally, we identify some open issues to motivate future research.

\section{Contrasting FFC and NFC Systems}
\begin{figure*}[!t]
\centering
\includegraphics[width=14.1cm]{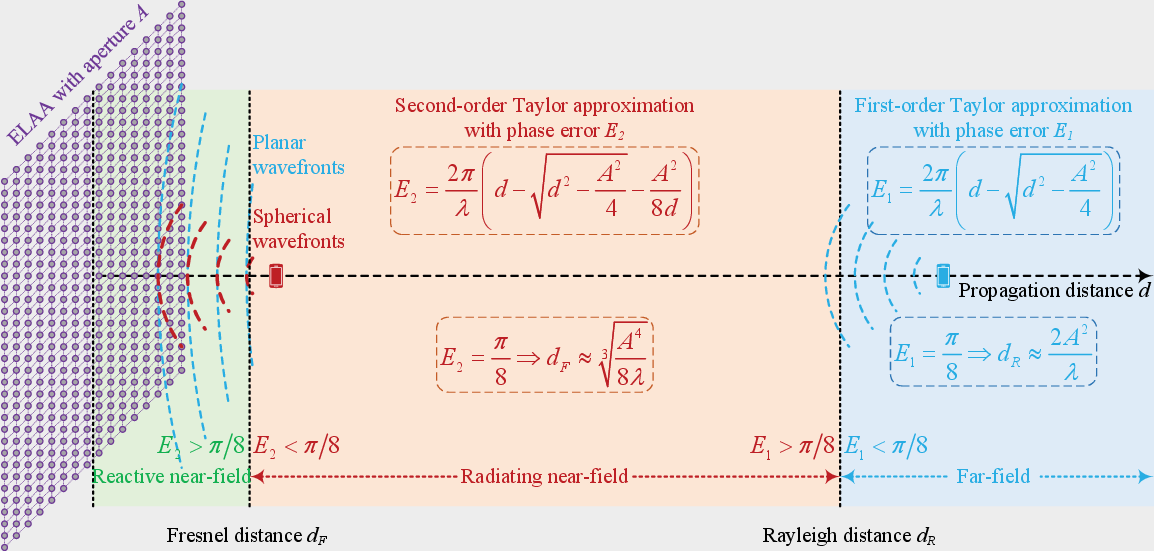}
\caption{Illustration of the far-field and near-field regions of an ELAA. The Fresnel distance and Rayleigh distance are determined by evaluating the phase errors caused by the parabolic and linear approximations of the spherical wavefronts, respectively.}\vspace{-0.6cm}
\label{fig_1}
\end{figure*}
As illustrated in Fig. \ref{fig_1}, the electromagnetic (EM) radiating field surrounding an ELAA can be roughly divided into the far-field and near-field regions. In reality, the transition between these two regions is gradual without a clear boundary. Nevertheless, a commonly used rule-of-thumb to distinguish the far- and near-field regions is the Rayleigh distance in terms of phase discrepancy across the array, also known as the Fraunhofer distance \cite{CM_2023_Cui_Near_56, arXiv_2023_Liu_Near_2}.

For users in the far-field region, the EM wavefronts arriving at the ELAA can be represented by the first-order Taylor (\emph{linear}) approximation of spherical wavefronts. All AEs experience roughly equal path loss and angle of arrival (AoA), leading to a uniform plane wave model. However, the linear approximation results in non-negligible phase error as the propagation distance decreases or the array aperture increases. In this context, the Rayleigh distance is defined as the minimum propagation distance at which the phase error across the array \emph{w.r.t.} a linear approximation is less than $\pi/8$, as shown in Fig. \ref{fig_1}. When the propagation distance is shorter than the Rayleigh distance, the user falls into the ELAA's near-field region. Hence, the associated spherical wavefronts must be considered for accurately characterizing the signal characteristics' variations across the ELAA \cite{TCOM_2022_Cui_Channel_107}. Observe from Fig. \ref{fig_1} that the near-field region can be further split into the reactive and radiating near-field (Fresnel) regions separated by the Fresnel distance. It is defined as the distance where the phase error caused by the second-order Taylor (\emph{parabolic}) -- rather than by the linear -- approximation is less than $\pi/8$ \cite{TWC_2022_Zhang_Beam_73}. Within the reactive near-field region, evanescent waves dominate and decay exponentially with distance. Since the reactive near-field region is typically covered by a few wavelengths, wireless communications mainly focus on the radiating near-field region, referred to as the near-field in this article for brevity.

In addition to using the phase discrepancy across the ELAA to separate the far- and near-field regions, \emph{Lu and Zeng} \cite{TWC_2022_Lu_Communicating} proposed the uniform-power distance (UPD) criterion to determine the boundary (\textbf{B}) based on channel gain variations. Within the UPD, the channel gain variations across the entire array are no longer negligible. Moreover, \emph{Cui et al.} \cite{CM_2023_Cui_Near_56} quantified the Rayleigh distance for reconfigurable intelligent surface (RIS)-based communication systems. As a further advance, \emph{Hu et al.}\cite{TWC_2023_Hu_Design} derived the Rayleigh distance to determine the Fresnel region, taking into account the off-boresight configuration to characterize the effects of signal directions on the maximum phase differences.

Compared to the distance-agnostic far-field model that assumes planar wavefronts based only on the AoA, the near-field model accounts for spherical wavefronts across the ELAA, which depend on both the propagation direction and distance. This focal position awareness allows NFC to gain benefits in three key aspects:
\begin{itemize}
\item \textbf{\emph{Spatial Multiplexing:}} Near-field MIMO using ELAAs exhibits substantial spatial multiplexing gains in LoS propagation conditions. This is in sharp contrast to far-field MIMO, which critically relies on rich scattering environments.
\item \textbf{\emph{Multiple Access:}} Spherical wavefronts allow beams to be focused in both angle and distance. This facilitates supporting multiple users even for those in the same angular direction relative to the ELAA.
\item \textbf{\emph{Wireless Sensing:}} Wider bandwidth and larger array aperture can significantly improve the positioning performance of 6G networks, potentially achieving centimeter-level accuracy. This enhanced capability can better support sensing the velocity, orientation, and other geometric information of targets of interest.
\end{itemize}

\section{Challenges and Recent Advances in NFC}
\begin{figure*}[!t]
\centering
\includegraphics[width=18.1cm]{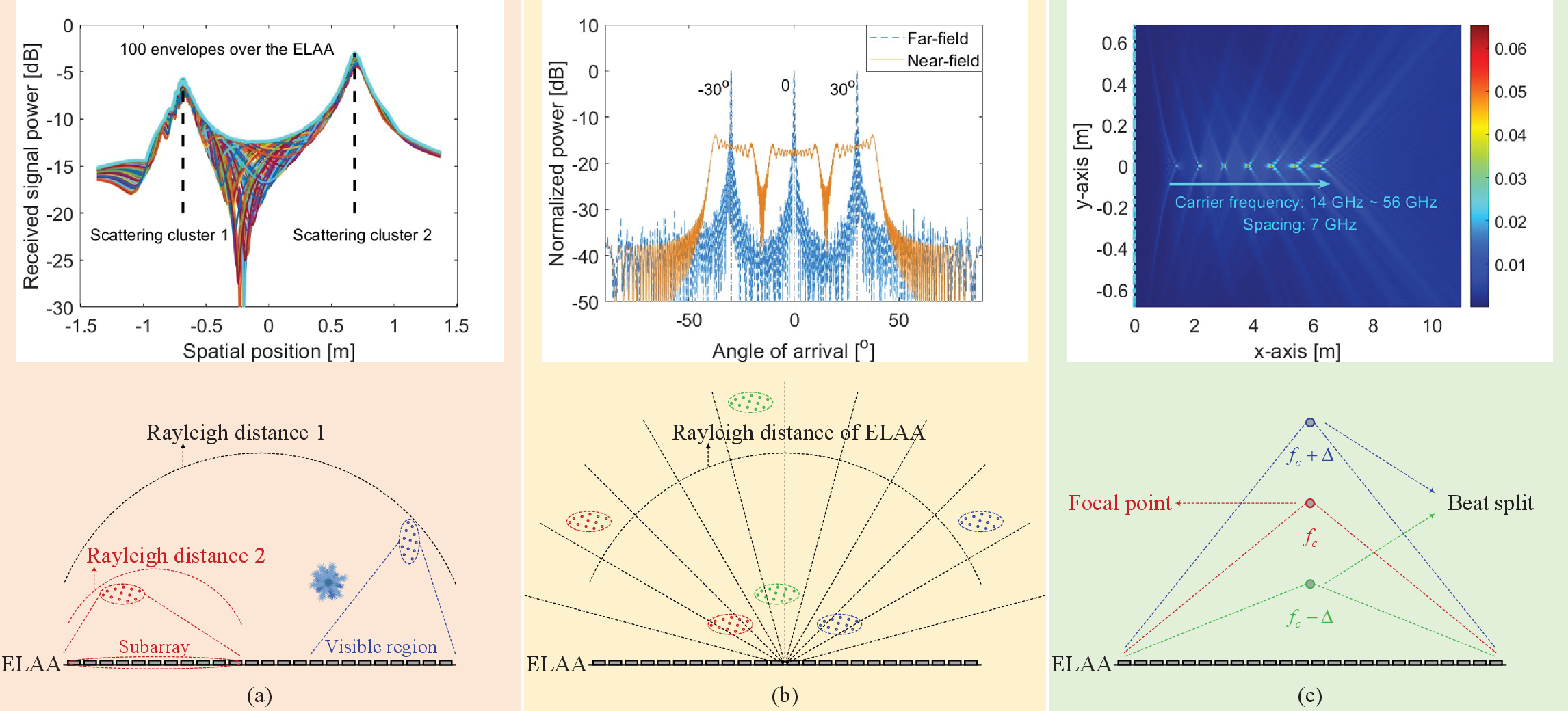}
\caption{(a) The spatial non-stationarity of the received signal strength, where we consider a ULA with an aperture of $2.74$ m operating at $28$ GHz with two scattering clusters; (b) The angular-domain spread of near-field channels, where we consider three scatterers that are $3$ m away from the ULA and have AoAs of $-30^{\circ}$, $0$, $30^{\circ}$, respectively; (c) The beam-split effect, where we examine a wideband NFC system with a ULA of length $2.74$ m.}\vspace{-0.6cm}
\label{fig_2}
\end{figure*}
In this section, we examine several key challenges in NFC system design. We also discuss the latest advances in near-field channel modeling and estimation as well as the developments of beamfocusing schemes and hardware architectures.
\subsection{Near-Field Channel Modeling (\textbf{CM})}
\textbf{\emph{Challenges:}} Due to the large antenna array (AA) aperture, different AEs experience noticeably different propagation distances and angles relative to the scatterers/users, hence resulting in distinct signal amplitude/phase variations across the entire ELAA. Additionally, practical communication scenarios generally involve multipath scattering environments, which are much more complex in NFC for several reasons. \emph{Firstly}, as shown in Fig. \ref{fig_2}(a), some scatterers/users are far away from the ELAA, while others are within its near-field region; \emph{Secondly}, the large array aperture means that its subarrays may observe distinct environments having different obstacles and scatterer clusters. \emph{Thirdly}, for ultra-wideband communications, low-frequency components having smaller Rayleigh distance may operate in the far-field, while high frequencies are in the near-field. These unique channel characteristics imply that the received signal power may vary dramatically across the array, as illustrated in Fig. \ref{fig_2}(a). An appropriate statistical model should accurately characterize both the spatial non-stationarity and near-field effects \cite{CM_2023_Cui_Near_56}.

\textbf{\emph{Advances:}} To simultaneously account for both the angular and distance information, \emph{Cui and Dai} \cite{TCOM_2022_Cui_Channel_107} proposed a polar-domain representation of the near-field channels. They utilized the Fresnel approximation of the near-field channel to design the Cartesian-to-polar transform matrix. Further, they theoretically demonstrated that only the non-uniform sampling along the distance dimension can achieve the minimum similarity among codewords in the transmit precoding (TPC) dictionary. Moreover, by explicitly modeling the physical size of each AE, \emph{Lu and Zeng} \cite{TWC_2022_Lu_Communicating} put forward a generic NFC channel model that took into account the variations of signal phase, amplitude, and projected aperture across the ELAA caused by different AoAs and propagation distances. Given this NFC model, they obtained a closed-form expression of the received signal-to-noise ratio (SNR) relying on the optimal single-user maximum ratio transmission (MRT), which reveals saturating SNR gains upon increasing the AA aperture. Recently, \emph{Liu et al.} \cite{arXiv_2023_Liu_Near_2} developed a general model for accurately capturing the free-space path loss and the signal loss caused by the effective array aperture and polarization mismatch due to the angular variations across the ELAA.

\subsection{Near-Field Channel Estimation (\textbf{CE})} 
\textbf{\emph{Challenges:}} Employing ELAAs for creating focal beams crucially depends on having accurate near-field channel state information (CSI), which is much more complex to obtain due to the huge number of AEs involved in ELAA systems. While channel sparsity due to the limited number of propagation paths at THz frequencies lends itself to compressed sensing (CS)-based channel estimation, the selection of dictionaries is particularly critical. Specifically, the classic Fourier dictionary \cite{TCOM_2022_Cui_Channel_107} provides a satisfactory sparse representation of far-field channels in the angle domain, as shown in Fig. \ref{fig_2}(b). Nevertheless, for NFC scenarios, the channel response depends on both the AoA and propagation distance. Observe from Fig. \ref{fig_2}(b) that a single near-field path component spans across multiple codewords in the Fourier dictionary. This inevitably degrades the performance of existing CS-based channel estimation techniques that rely on angular sparsity. As a result, an appropriate TPC codebook and CS algorithm design is required for accurately estimating the near-field channels at a low pilot overhead.

\textbf{\emph{Advances:}} In \cite{TCOM_2022_Cui_Channel_107}, \emph{Cui and Dai} demonstrated the sparsity of near-field channels in terms of polar -- rather than Cartesian -- coordinates. Based on this, they utilized the popular simultaneous orthogonal matching pursuit and the simultaneous iterative gridless weighted algorithms to estimate the near-field channels for on-grid and off-grid scenarios, respectively. In the off-grid case, the channel parameters such as path gains, angles, and distances are directly estimated. They validated that in the near-field regions, these polar-domain channel estimation schemes achieve enhanced mean square error (MSE) performance compared to existing approaches. Motivated by the low-dimensional characteristic of parametric channel estimation, \emph{Dardari et al.} \cite{TWC_2022_Dardari_LOS} developed a single-anchor localization technique for mobile users in the near-field regime of a RIS. They proposed two practical signaling and positioning methods considering the orthogonal frequency division multiplexing (OFDM) downlink. Near-field propagation enhances the robustness of localization even when stochastic obstructions occur between the RIS and the user. Additionally, \emph{Elbir et al.} \cite{OJCS_2023_Elbir_Elbir_7} conceived channel estimation for ultra-wideband systems for counteracting the beam-split effect. Specifically, a sparse Bayesian learning (SBL) technique was developed for joint channel and beam-split estimation. The LoS-dominated nature of the THz channel was leveraged for reducing the computational complexity.

\begin{figure*}[!t]
\centering
\includegraphics[width=18.1cm]{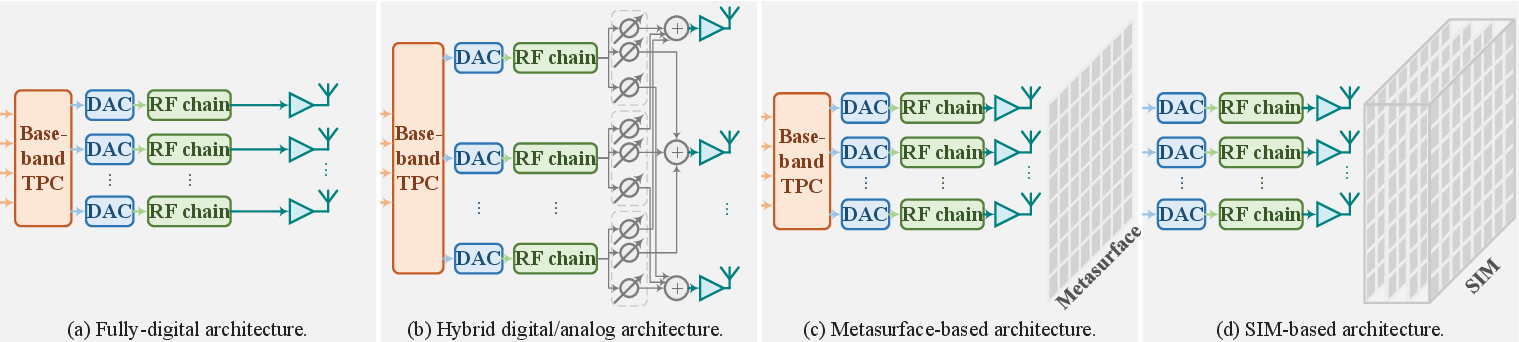}
\caption{Four types of hardware architectures for implementing ELAAs.}\vspace{-0.6cm}
\label{fig_3}
\end{figure*}
\subsection{Near-Field Beamfocusing (\textbf{BF})}
\textbf{\emph{Challenges:}} As mentioned earlier, by leveraging spherical wavefronts near-field beamfocusing concentrates the radiated energy towards a specific focal point, which requires accurate tuning of the TPC weights associated with each AE and data stream. Additionally, to enable a cost-efficient implementation, the popular hybrid architecture using analog phase-shifters is widely adopted. However, in wideband NFC systems, the hybrid architecture applies the same phase-shifts across the operating frequency band. This results in the detrimental ``beam-split'' phenomenon where the beams generated by the analog beamformer diverge at different subcarrier frequencies, as shown in Fig. \ref{fig_2}(c). Explicitly, beam-split limits the effective bandwidth in ELAA systems and severely degrades the received signal energy at the intended focal points. Integrating time-delay architectures with phase-shifters can realize frequency-selective beamfocusing at the cost of extra implementation complexity and power consumption. Therefore, developing efficient signal processing techniques for mitigating the beam-split effect at affordable hardware costs presents significant challenges for designing ELAAs.

\textbf{\emph{Advances:}} Next, we will review the latest technological advances in near-field beamfocusing, covering point-to-point, multiuser, and wideband systems.
\begin{itemize}
\item \emph{Point-to-Point Scenario (\textbf{P2P}):} Based on electromagnetic theory, \emph{Dardari} \cite{JSAC_2020_Dardari_Communicating_133} analyzed the optimal communication modes in NFC systems and derived closed-form expressions for both the link gain and spatial DoF. He demonstrated that upon using a large intelligent surface (LIS), the available DoF can be higher than $1$ even in strong LoS conditions, resulting in a significant increase in spatial multiplexing gain. This is in contrast to conventional MIMO-FFC systems that can only provide beamforming gain in LoS conditions. Recently, \emph{Hu et al.} \cite{TWC_2023_Hu_Design} proved that the optimal 3D near-field beamfocusing can be decomposed into a 2D far-field beamformer and a one-dimensional (1D) near-field beamfocuser in the Fresnel region. The former is used to compensate for phase variations caused by off-boresight configuration, while the latter compensates for the remaining phase variations due to distance differences across the LIS elements. Such a two-phase beamfocusing design significantly reduces the TPC codebook size because a pair of small codebooks facilitate the creation of a large composite codebook by hierarchically combining their codewords. They may also be backward compatible with the existing 5G new radio (NR) protocol.
\item \emph{Multiuser Scenario (\textbf{MU}):} To examine the beamfocusing capability of ELAAs, \emph{Zhang et al.} \cite{TWC_2022_Zhang_Beam_73} investigated the MIMO-NFC downlink employing both fully digital arrays, hybrid analog/digital arrays, and dynamic metasurface antennas (DMAs). They demonstrated that the beamfocusing technique substantially improved the achievable sum rate of the MIMO-NFC networks. Recently, \emph{Wei et al.} \cite{TWC_2023_Wei_Tri} utilized triple-polarization patch antennas for improving the capacity and reliability of multi-user holographic MIMO systems operating in the NFC regime. A pair of TPC schemes were developed for mitigating both cross-polarization and inter-user interference. In \cite{JSAC_2023_Wu_Multiple}, \emph{Wu and Dai} proposed location division multiple access (LDMA) leveraging the focusing capability of near-field beams to serve multiple users located at different positions without significant interference. They proved that near-field beamfocusing vectors tend to be asymptotically orthogonal in the distance domain as the size of ELAA increases. Hybrid precoding was utilized to focus signals on specific locations with minimum leakage, thus substantially improving spectrum efficiency.
\item \emph{Wideband Scenario (\textbf{WB}):} In \cite{TWC_2022_Myers_InFocus_20}, \emph{Myers and Heath} evaluated the ability of a phased-array-based circular planar array (CPA) to focus a beam in wideband NFC with the receiver located at the AA's boresight. They determined the array size, propagation distance, and bandwidth that resulted in a non-negligible beam-split. To make the beamfocuser more robust against near-field beam-split, a frequency-modulated continuous waveform (FMCW) chirp sequence was designed along the transmit array's dimension to control phase-shifts. They demonstrated that the beams generated achieve approximately constant beamforming gain over a wide bandwidth. By contrast, upon exploiting the near-field beam-split effect, \emph{Cui and Dai} \cite{TWC_2023_Cui_Near} designed an efficient wideband beam training scheme having low pilot overhead. Specifically, they employed a time-delay architecture for generating multiple beams focusing on multiple locations within a given distance range in each time slot. In different time slots, different distance ranges are probed by fine-tuning the time-delay parameters. Thus, the angle and distance are searched in frequency-division and time-division manners, respectively. In \cite{OJCS_2023_Elbir_Elbir_7}, \emph{Elbir et al.} addressed the beam-split effect by modeling it as an array perturbation.
\end{itemize}

\subsection{Hardware Architecture (\textbf{HA})}
\begin{table*}[!t]
\doublespacing
\caption{A survey of recent advances in near-field communications.}
\label{tab1}
\centering
\footnotesize
\begin{tabular}{cc|c|c|c|l}
\hline
\multicolumn{2}{c|}{Characteristics} & Ref. & Array & Channel & Main contributions \\ \hline
\multicolumn{1}{c|}{\multirow{4}{*}{\textbf{B}}} & \multicolumn{1}{l|}{\multirow{2}{*}{Phase}} & \cite{CM_2023_Cui_Near_56} & UPA & LoS & Determined the Rayleigh distance for MIMO and RIS-assisted communication systems.\\ \cline{3-6} 
\multicolumn{1}{c|}{} & \multicolumn{1}{l|}{} & \cite{TWC_2023_Hu_Design} & UPA & LoS & Analyzed the Rayleigh distance considering the off-boresight setup. \\ \cline{2-6} 
\multicolumn{1}{c|}{} & \multicolumn{1}{l|}{Amplitude} & \cite{TWC_2022_Lu_Communicating} & ULA/UPA & LoS & Defined the UPD concerning the non-negligible signal amplitude variations across the ELAA. \\ \cline{2-6} 
\multicolumn{1}{c|}{} & \multicolumn{1}{l|}{DoF} & $\star$ & ULA & LoS & Analyzed the impact of different propagation distances on the spatial DoF. \\ \hline
\multicolumn{2}{c|}{\multirow{3}{*}{\textbf{CM}}} & \cite{arXiv_2023_Liu_Near_2} & ULA/UPA & Multipath & Proposed an NFC channel model considering effective aperture and polarization mismatch. \\ \cline{3-6} 
\multicolumn{2}{c|}{} & \cite{TWC_2022_Lu_Communicating} & ULA/UPA & LoS & Proposed a unified NFC model to characterize the phase/amplitude variations and aperture loss. \\ \cline{3-6} 
\multicolumn{2}{c|}{} & \cite{TCOM_2022_Cui_Channel_107} & ULA & Multipath & Proposed a polar coordinate representation of the near-field channels. \\ \hline
\multicolumn{2}{c|}{\multirow{3}{*}{\textbf{CE}}} & \cite{OJCS_2023_Elbir_Elbir_7} & ULA & Multipath & Put forward an SBL-based approach for jointly estimating the THz channel and beam-split.\\ \cline{3-6}
\multicolumn{2}{c|}{} & \cite{TWC_2022_Dardari_LOS} & UPA & Multipath & Leveraged a single anchor and a RIS to probe the localization of mobile users. \\ \cline{3-6}
\multicolumn{2}{c|}{} & \cite{TCOM_2022_Cui_Channel_107} & ULA & Multipath & Estimated the near-field channels by exploiting the channel sparsity in the polar domain. \\ \hline
\multicolumn{1}{c|}{\multirow{6}{*}{\textbf{BF}}} & \multirow{2}{*}{\emph{P2P}} & \cite{TWC_2023_Hu_Design} & UPA & LoS & Proposed a ``2D + 1D'' near-field beamfocusing design to compensate for phase variations. \\ \cline{3-6} 
\multicolumn{1}{c|}{} & & \cite{JSAC_2020_Dardari_Communicating_133} & Continuous & LoS & Derived analytical expressions for the link gain and spatial DoFs based on electromagnetism. \\ \cline{2-6} 
\multicolumn{1}{c|}{} & \multirow{3}{*}{\emph{MU}} & \cite{TWC_2022_Zhang_Beam_73} & UPA & LoS & Evaluated the beamfocusing ability of fully digital, hybrid, and DMA architectures. \\ \cline{3-6} 
\multicolumn{1}{c|}{} & & \cite{TWC_2023_Wei_Tri} & UPA & LoS & Utilized the triple polarization for multi-user holographic MIMO communication systems. \\ \cline{3-6}
\multicolumn{1}{c|}{} & & \cite{JSAC_2023_Wu_Multiple} & ULA/UPA & Multipath & Proposed LDMA to simultaneously serve multiple users in the near-field region. \\ \cline{2-6} 
\multicolumn{1}{c|}{} & \multirow{2}{*}{\emph{WB}} & \cite{TWC_2022_Myers_InFocus_20} & CPA & LoS & Proposed a spatial FMCW chirp-based beamfocusing approach to mitigate the beam-split effect. \\ \cline{3-6} 
\multicolumn{1}{c|}{} & & \cite{TWC_2023_Cui_Near} & ULA & LoS & Proposed a fast beam training scheme leveraging the near-field beam-split effect. \\ \hline
\multicolumn{2}{c|}{\multirow{2}{*}{\textbf{HA}}} & \cite{TWC_2022_Zhang_Beam_73} & UPA & LoS & Proposed a DMA architecture with densely-spaced meta-atoms to realize ELAAs. \\ \cline{3-6} 
\multicolumn{2}{c|}{} & \cite{JSAC_2023_An_Stacked} & UPA & Multipath & Proposed an SIM-based transceiver architecture to accomplish TPC in the wave domain. \\ \hline
\end{tabular}\vspace{-0.6cm}
\end{table*}
\textbf{\emph{Challenges:}}
Designing appropriate hardware architectures for implementing efficient ELAAs that operate at high frequencies is challenging for several reasons. \emph{Firstly}, the efficiency of a power amplifier typically decreases as bandwidth increases. \emph{Secondly}, as the number of active AEs becomes extremely large, fully digital architectures (see Fig. \ref{fig_3}(a)) that assign each AE a dedicated RF chain become excessively expensive and power-thirsty. \emph{Thirdly}, the resultant computational complexity of beamfocusing increases accordingly. To address these issues, a cost-efficient option is the hybrid architecture that uses a limited number of RF chains for low-dimensional digital beamforming, followed by a set of low-cost analog phase-shifters, as shown in Fig. \ref{fig_3}(b). However, due to the limited resolution of the phase-shifters, the hybrid beamformer suffers from \emph{i)} reduced beamfocusing gain; \emph{ii)} inability of having multiple focal points; and \emph{iii)} beam-split effects due to the inherent frequency-flat profile of phase-shifters. Additionally, the power consumption of a large number of phase-shifters is still considerable. Therefore, it is crucial to design cost- and power-efficient hardware architectures that are scalable for ELAAs and high frequencies.

\textbf{\emph{Advances:}} Recently, advanced metasurface technology has been developed for the programmable amelioration of EM behavior by ELAAs at a low cost and power consumption (see Fig. \ref{fig_3}(c)). In \cite{TWC_2022_Zhang_Beam_73}, the DMA architecture was fabricated using multiple microstrips, each containing many subwavelength-spaced radiating metamaterial elements. The frequency response of each element may be externally adjusted by varying its local electrical properties. However, the tuning precision of the metasurface structure was still constrained by the discrete components. To further simplify the structure, the authors of \cite{JSAC_2023_An_Stacked} proposed an intriguing \emph{stacked intelligent metasurfaces (SIM)}-based transceiver architecture for ELAA, where the digital TPC was eliminated. As shown in Fig. \ref{fig_3}(d), SIM is capable of forming a MIMO TPC and a receiver combiner, as the EM waves propagate through it. The revolutionary SIM technology offers at least three benefits. \emph{Firstly}, SIM has the potential of approaching the performance of the digital TPC by mimicking a neural network architecture using physical metasurfaces. \emph{Secondly and most remarkably}, the forward propagation within the SIM occurs at the speed of light, which significantly reduces signal processing latency. \emph{Thirdly}, each spatial stream is radiated and recovered independently at the corresponding transmit and receive ports, inherently reducing the number of RF chains and signal processing complexity.

The recent technological advances in NFC are summarized in Table \ref{tab1} in a nutshell.
\section{Case Studies}
In this section, we present two case studies for demonstrating the potential of NFC in improving the spatial DoF and enhancing positioning performance.
\subsection{Improved Spatial Multiplexing Gain}
\begin{figure*}[!t]
\centering
\includegraphics[width=18.1cm]{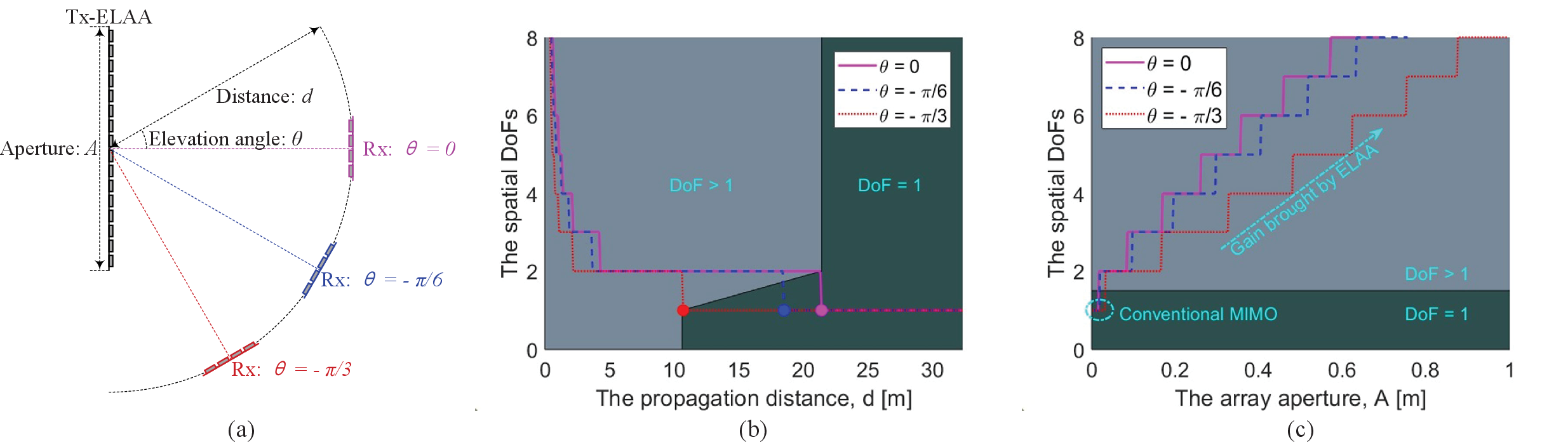}
\caption{Point-to-point MIMO-NFC under the LoS propagation: (a) The system layout; (b) The spatial DoFs versus the propagation distance; (c) The spatial DoFs versus the array aperture.}\vspace{-0.6cm}
\label{fig_4}
\end{figure*}

We first validate the DoF increase in near-field point-to-point MIMO channels under LoS propagation conditions. The specific AA layout is shown in Fig. \ref{fig_4}(a). Both the transmitter (TX) and receiver (RX) are equipped with a ULA, having half-wavelength spacing between AEs. The system operates at a carrier frequency of $300$ GHz. The RX has $16$ antennas, corresponding to an aperture of $0.75$ cm. In our simulations, we consider both on-boresight and off-boresight configurations, where the RX is located at angles of $0$, $\pi/6$, and $\pi/3$ from the normal direction of the TX array.

Fig. \ref{fig_4}(b) shows the effects of propagation distance on the spatial DoF. An ELAA having $1,024$ antennas spanning $A=0.51$ m is deployed at the TX. The DoFs are computed allowing only for 1\% total energy loss. Observe from Fig. \ref{fig_4}(b) that the spatial DoFs increase from $1$ to $8$ as the propagation distance decreases from $30$ m to $0.5$ m. Thanks to the higher DoFs, the near-field distance-aware channel can simultaneously transmit multiple data streams using appropriate MIMO TPC, giving rise to a significant increase in channel capacity. For off-boresight setups, Fig. \ref{fig_4}(b) shows that a shorter propagation distance is required to get an improved DoF, as the elevation angle increases. Additionally, in the setup considered, the Rayleigh distance is $538$ m, while the rank boost occurs at $d = 21.3$ m. Therefore, new distance criteria based on changes in essential spatial DoF have to be developed. Moreover, Fig. \ref{fig_4}(c) shows the DoF versus the array aperture $A$. The propagation distance is set to $d = 0.7$ m. In conventional MIMO systems having an AA aperture smaller than $2$ cm, the DoF is $1$ under LoS conditions, providing only beamforming gain. As the AA aperture increases, the DoF substantially improves. For the ELAA larger than $0.9$ m, the spatial DoF is larger than $8$ for all setups considered. Again, non-parallel Tx-Rx arrangements yield achievable DoF lower than that obtained by parallel ULAs, obeying the optimal geometric configuration \cite{JSAC_2020_Dardari_Communicating_133}.

\subsection{Enhanced Positioning Accuracy}
Next, we evaluate the performance of positioning for a single user in the near-field region, at a carrier frequency of $300$ GHz. Specifically, we examine a setup with four access points (APs) located at $(3,0)$ m, $(6,3)$ m, $(3,6)$ m, $(0,3)$ m. Each AP is equipped with a ULA lined up with the coordinate axes, with half-wavelength AE spacing. The least squares (LS) estimator is employed to calculate the user's position based on the time-of-arrival (ToA) measurements of the signals received at the four APs. Each measurement error has a variance of $15$ dB m$^2$. We run 10,000 independent experiments to study the circular error probability (CEP), which is defined as the radius of a circle centered on the estimated position's mean, where half of the location estimates fall within the circle.

\begin{figure}[!t]
\centering
\subfloat[Far-field positioning.]{\includegraphics[width=4.4cm]{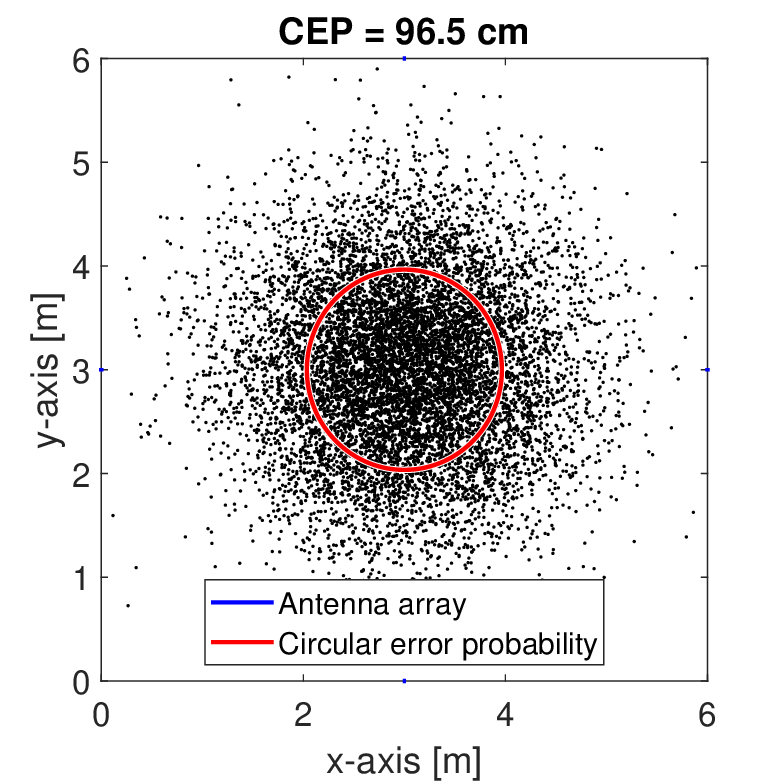}%
}
\subfloat[Near-field positioning.]{\includegraphics[width=4.4cm]{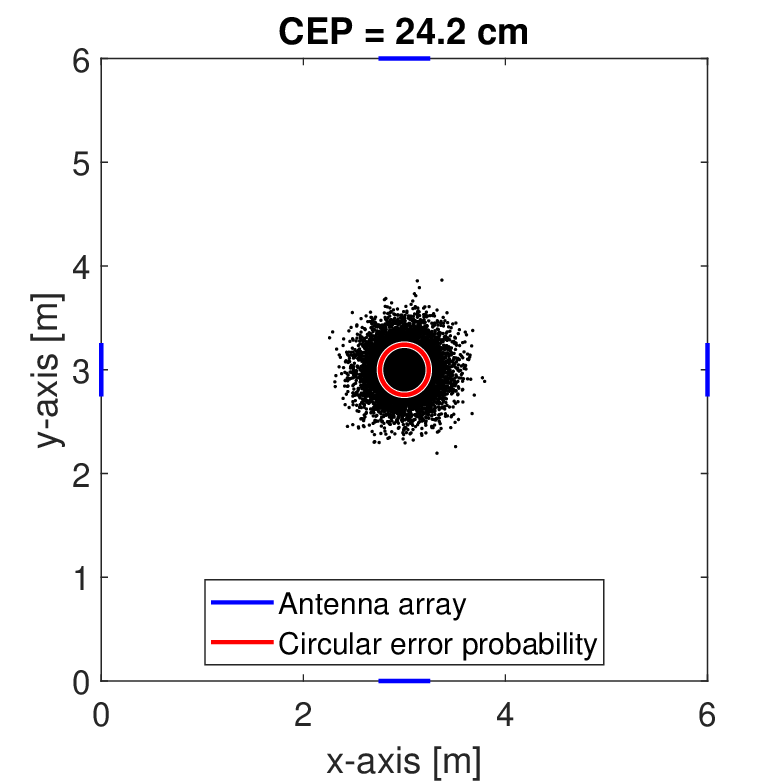}%
}
\caption{Contrasting near-field and far-field positioning performance.}\vspace{-0.6cm}
\label{fig_5}
\end{figure}

We first consider the case where four ULAs are each equipped with $64$ AEs, corresponding to a length of $3$ cm. Fig. \ref{fig_5}(a) shows the distributions of the estimated user position obtained by employing the LS method. The circle in the figure is centered at the actual user location of $\left ( 3,3 \right )$ m. The CEP in Fig. \ref{fig_5}(a) was found to be $96.5$ cm. When increasing the number of antennas to $1024$ for each AP, corresponding to an aperture of $0.51$ m, the user enters the near-field region of all arrays. As shown in Fig. \ref{fig_5}(b), the CEP using the LS estimator in the near-field is reduced to $24.2$ cm, outperforming its far-field counterpart by approximately $72.3$ cm. Further increasing the size of the AAs would result in even better positioning accuracy.

In addition to these typical examples we already discussed, NFC can also enhance both physical-layer security (PLS) and multi-user communication scenarios by taking advantage of the distance-aware spherical wavefronts \cite{arXiv_2023_Liu_Near_2, CM_2023_Cui_Near_56}, which provide significant research opportunities.
\section{Future Directions}
Despite its great potential, designing and implementing NFC still requires future innovations in some important issues, which provides rich opportunities for future research. This section identifies some potential research directions.
\subsection{Performance Evaluation}
It is crucial to evaluate the performance limit in the near-field regions in terms of various target objectives, e.g., the spatial DoF. Since the near-field behavior varies for the on-axis and off-axis configurations, further evaluation of both the angular and distance dependency is also required. The orientation of AAs also has a non-negligible effect on the near-field communication and sensing performance. Hence, developing an analytical DoF expression vs. the propagation distance and relative array placement is of significance for guiding practical NFC system design.
\subsection{Spatial Non-Stationarity}
Modeling spatial non-stationarity is still an open challenge, especially in complex near-field propagation environments. The variations in polarization and effective aperture across large AAs must be taken into account in the statistical NFC models. Again, the large array aperture implies that different subarrays of the ELAA may observe different scattering environments, resulting in spatial selectivity. How to leverage this spatial non-stationarity to design relevant transmission schemes warrants further study.
\subsection{SIM-Based Architecture}
While SIM can perform beamfocusing in the wave domain, its performance in wideband NFC systems is yet to be evaluated. A notable point is that the multilayer structure of SIM has a frequency-selective profile, which provides a new design DoF for optimizing wideband beamfocusing. Finally, the power consumption of compensating for the energy losses of propagating through multiple metasurfaces and those caused by conventional digital components has to be compared at the system level. A mathematically tractable and physically consistent model for characterizing the near-field propagation in SIM is urgently needed.
\subsection{Fine-Granularity Sensing}
Upon tapping into wider bandwidth and larger AAs, NFC is expected to provide more advanced perception capabilities, allowing for detecting fine-granularity geometric features of objects. For instance, subtle gestures can be discerned by detecting the Doppler shifts. Moreover, the millimeter-level positioning precision enables applications such as holographic imaging and 3D mapping. While NFC promises accurate sensing capacities over wireless networks, there are still major technical challenges that have to be addressed, such as developing unified protocols and waveform designs.

\section{Conclusions}
This article provided an overview of the potential and challenges of designing NFC systems. We discussed the key characteristics of NFC relying on spherical wavefronts and highlighted its difference \emph{w.r.t.} conventional FFC. Additionally, we identified some key technical challenges for NFC system design and reviewed recent advances in near-field channel modeling and estimation, beamfocusing schemes, and hardware architectures. Two case studies demonstrated the potential of NFC. Finally, we pointed out some directions for future development of NFC.

\section*{Acknowledgments}
This research is supported by the Ministry of Education, Singapore, under its MOE Tier 2 (Award number MOE-T2EP50220-0019) and by the Science and Engineering Research Council of A*STAR (Agency for Science, Technology, and Research) Singapore, under Grant No. M22L1b0110.

\bibliographystyle{IEEEtran}
\bibliography{ref}
\newpage
\section*{Biographies}
\vspace{-30pt}
\begin{IEEEbiographynophoto}{Jiancheng An}
is currently a Research Fellow with the Engineering Product Development (EPD) Pillar, Singapore University of Technology and Design, Singapore.
\end{IEEEbiographynophoto}
\vspace{-18pt}
\begin{IEEEbiographynophoto}{Chau Yuen}[F]
is currently an Associate Professor at the Nanyang Technological University, Singapore. He is an Editor of IEEE Transactions on Vehicular Technology and IEEE Transactions on Network Science and Engineering.
\end{IEEEbiographynophoto}
\vspace{-18pt}
\begin{IEEEbiographynophoto}{Linglong Dai}[F]
is a professor at Tsinghua University, China. His current research interests include massive MIMO, RIS, wireless AI, and EIT. He was listed as a Highly Cited Researcher by Clarivate from 2020 to 2022.
\end{IEEEbiographynophoto}
\vspace{-18pt}
\begin{IEEEbiographynophoto}{Marco Di Renzo}[F]
is a CNRS Research Director (Professor) and the Head of Intelligent Physical Communications in the Laboratory of Signals and Systems at Paris-Saclay University CNRS and CentraleSupelec.
\end{IEEEbiographynophoto}
\vspace{-18pt}
\begin{IEEEbiographynophoto}{M\'erouane Debbah}[F]
is with Khalifa University of Science and Technology, Abu Dhabi, UAE.
\end{IEEEbiographynophoto}
\vspace{-18pt}
\begin{IEEEbiographynophoto}{Lajos Hanzo}[F]
received Honorary Doctorates from the Technical University of Budapest and Edinburgh University. He is a Foreign Member of the Hungarian Science Academy, a Fellow of the Royal Academy of Engineering (FREng), of the IET, of EURASIP, and holds the IEEE Eric Sumner Technical Field Award.
\end{IEEEbiographynophoto}
\vfill
\end{document}